\documentclass[A4paper,preprint,12pt,tightenlines,amsmath,amssymb,floatfix,nofootinbib,aps,raggedbottom]{revtex4-2}

\usepackage{graphicx}
\usepackage{bm}
\usepackage{url,multirow,xcolor}
\usepackage[font=footnotesize,margin=24pt,justification=centerlast]{caption}

\begin{document}

\title{ Half-life of $^{\bm{136}}$Xe for neutrinoless double-$\bm{\beta}$ decay calculated with effective axial-vector current coupling unified for two-neutrino and neutrinoless double-$\bm{\beta}$ decay modes
\vspace{5pt} }

\author{J.\ Terasaki \\ \vspace{0pt}}
\affiliation{ Institute of Experimental and Applied Physics\hbox{,} Czech Technical University in Prague, Husova 240/5, 110\hspace{3pt}00 Prague 1, Czech Republic}

\author{O.\ Civitarese \\ \vspace{0pt}}
\affiliation{\raisebox{0pt}{Department of Physics, University of  La Plata, 49 y 115.~C.C.~67 (1900)}, La Plata, Argentina \\
and \hbox{IFLP-CONICET, diag 115 y 64. La Plata, Argentina}\vspace{10pt}}


\begin{abstract} 
The upper limit on the mass of the Majorana neutrino, extracted from the limits on the nonobservation of the neutrinoless double-$\beta$ ($0\nu\beta\beta$) decay, is hampered by uncertainties in the matrix elements of the transition operators. Recently, we have shown  that the values of the effective axial-vector current coupling constants ($g_A^\textrm{eff}$) for the $0\nu\beta\beta$ and the two-neutrino double-$\beta$ decays are close. This striking result was obtained for the first time by including vertex corrections and two-body currents in these matrix elements. In this letter, we calculate the half-life for the $0\nu\beta\beta$ decay ($T_{1/2}^{0\nu}$) of $^{136}$Xe using this closeness and show the convergence of the half-life with respect to the variation of the method to determine $g_A^\textrm{eff}$. The closeness of the $g_A^\textrm{eff}$ of the two decay modes plays a decisive role in predicting $T_{1/2}^{0\nu}$. 
The appropriate value of $g_A^\textrm{eff}$ depends on the  assumptions made for the sectors of the nuclear structure and transition operators of the calculations within  the perturbation scheme. 
The value $g_A^\textrm{eff}\approx 1$ is obtained  when the SkM$^\ast$ is used to describe the nuclear structure component, while a smaller value of  $g_A^\textrm{eff}$ is obtained by applying a less realistic interaction like the SGII one. 
\end{abstract}


%
\maketitle
\newpage
%

The accurate prediction of the half-life for the neutrinoless double-$\beta$ ($0\nu\beta\beta$) decay ($T_{1/2}^{0\nu}$) is a highly demanded task. It has called the attention of experimentalists and theorists for decades \cite{Pri81,Suh98,Ago23}, and it is in our days a matter of intense activity in both fields. The observation of the $0\nu\beta\beta$ decay is the evidence of the existence of the Majorana neutrino \cite{Doi85}, and this observation indicates new physics beyond the standard model at least because of the violation of the lepton number conservation. It is also possible to discuss a contribution of the right-handed neutrino to the $0\nu\beta\beta$ decay. A strong foundation would be given to leptogenesis \cite{Fuk86,Gom23}, and the impact of such a discovery would reflect upon new developments in particle physics, astroparticle physics, and cosmology.

The calculation of $T_{1/2}^{0\nu}$ of the $0\nu\beta\beta$ decay requires the knowledge of four factors: i) the nuclear matrix elements (NME) of the operator acting on the nucleus in the decay, ii) the coupling constant of the axial-vector current and a nucleon ($g_A$), iii) the electron contribution (the phase space factor), and iv) the value of the effective mass of the Majorana neutrino $\langle m_\nu\rangle$ \cite{Doi85}. In this article, we assume only the light-mass neutrino. For the last factor, only weak constraints on the relation of $\langle m_\nu \rangle$ and the eigen neutrino masses are known. $T_{1/2}^{0\nu}$ can be calculated if a value of $\langle m_\nu \rangle$ is assumed. If $T_{1/2}^{0\nu}$ is measured, inversely, $\langle m_\nu \rangle$ can be deduced. The main difficulty in the relation between $T_{1/2}^{0\nu}$ and $\langle m_\nu\rangle$ is due to the uncertainty which originates in the theoretical-method dependence of the NME. Among the approximations for the nuclear wave functions used in the NME calculations are the shell model \cite{Men18,Hor13,Cor20b,Cor24}, the ab initio approach \cite{Nov21,Bel23,Bel24}, the quasiparticle random-phase approximation (QRPA) \cite{Hyv15,Sim18b,Fan18,Ter20}, the interacting boson model \cite{Bar15}, and the energy density functional approach \cite{Rod10,Vaq13,Son17}. 

The systematics of NME, for both the two-neutrino double-$\beta$ ($2\nu\beta\beta$) and the $0\nu\beta\beta$ decay, focuses on the sensitivity of the calculations upon the nuclear interactions and on the methods 
employed to calculate the participant nuclear wave functions. 
The literatures are rich concerning these aspects. Among the articles already published we may quote some of the recent ones which are closer to the subject of our present effort.

To start with, the dependence of the results upon the interactions was studied in Refs.~\cite{Hin22,Lv23}, where the known dependence upon the strength of the isovector proton-neutron interaction, already advanced many years ago \cite{Civ87}, was revisited for the  various Skyrme interactions in the context of the QRPA. Naturally, the aim of these studies was to fix the strength of the isovector proton-neutron interaction from the observed two-neutrino mode in order to reduce the uncertainties in the model dependence of the matrix elements of the neutrinoless mode.

Another attempt was reported in Ref.~\cite{Nov21}, where the coupled-channel calculations with the chiral effective field theory were performed for the $0\nu\beta\beta$ decay mode. The  results of such an approach yield smaller values of NME. In that work, the authors have explored also the quenching factor induced by the inclusion of two-body currents. We shall return to this point later on in the present work.

A similar approach within the chiral effective field theory, including the ab-initio uncertainty quantification of the NME for the $0\nu\beta\beta$ decay mode, was reported more recently \cite{Bel24}. Along this line, the work of Ref.~\cite{Bra22} focused on the calculation of the double-Gamow-Teller (double-GT) states within the framework of an effective field theory in order to determine the low-energy constants of the theory. 

Another approach, based on the use of the nuclear density functional theory, was reported in Ref.~\cite{Hin22}. The main contribution of the just quoted reference was the analysis of the obtained results for the half-lives of single and double $\beta$ decay transitions. The systematics reported in Ref.~\cite{Hin22} included measured and unmeasured  decay channels of the $\beta\beta$ decay transitions calculated in a novel method.

Concerning the efforts related to the calculation of the single and the double $\beta$ decays in light nuclei,  we shall mention the results reported in Ref.~\cite{Bas20} for the $0\nu\beta\beta$ decay in very light-mass nuclei. The theoretical approach of Ref.~\cite{Bas20}, based on the renormalization group theory, shows the importance of  performing cross-checks by the ab-initio methodology. 

Last but not least, we shall quote the shell model calculations reported in Ref.~\cite{Cor20b}.  The procedure started from a realistic two-body interaction, and it yielded an effective shell model Hamiltonian. The prediction of the decay for several candidates was then extremely useful, particularly for shedding light on the contribution of the short-range correlations.


The NME and $T_{1/2}^{0\nu}$ obtained with these methods show a large dispersion \cite{Eng17,Ago23}, which reflects upon the estimates of the neutrino mass. It is usual for the known weak decays to use an effective $g_A$ ($g_A^\textrm{eff}$) to reproduce the half-life \cite{Bro85}, while the vector current coupling constant $g_V$ is always set to one. Concerning $g_A$, an unconfirmed speculation is possible that $g_A^\textrm{eff}$ for the $0\nu\beta\beta$ and the other weak decay modes are quite different because the Majorana neutrino in the $0\nu\beta\beta$ decay is a virtual particle. 

In our work, we are looking at the consistency of the results for both the $2\nu$ and $0\nu$ $\beta\beta$-decay modes, paying attention to the value of $g_A^\mathrm{eff}$.
The problem is of upmost importance at the time of making predictions for the still undetected neutrinoless mode of decay. 
In coincidence with the findings in other studies recently reported \cite{Rho23, Rho25}, we have found that the value of $g_A$ should be close to unity. In our case, we have obtained this remarkable result  from the analysis of the behaviour of the NME calculated for the decay of $^{136}$Xe with the Skyrme interaction in the framework of the QRPA. As we shall discuss later on, the results of the present work, obtained within the nuclear structure context, are similar to the findings reported in a very different context and independently of our approach \cite{Rho23, Rho25}. 
Obviously these results have an impact about the prediction of the order of magnitude of the half-life. We shall show that the reduction in the value of $g_A$ affects critically the estimation of $T^{0\nu}_{1/2}$.  

\begin{figure}[t]
\begin{minipage}[t]{0.9\linewidth}
\includegraphics[width=1.0\textwidth]{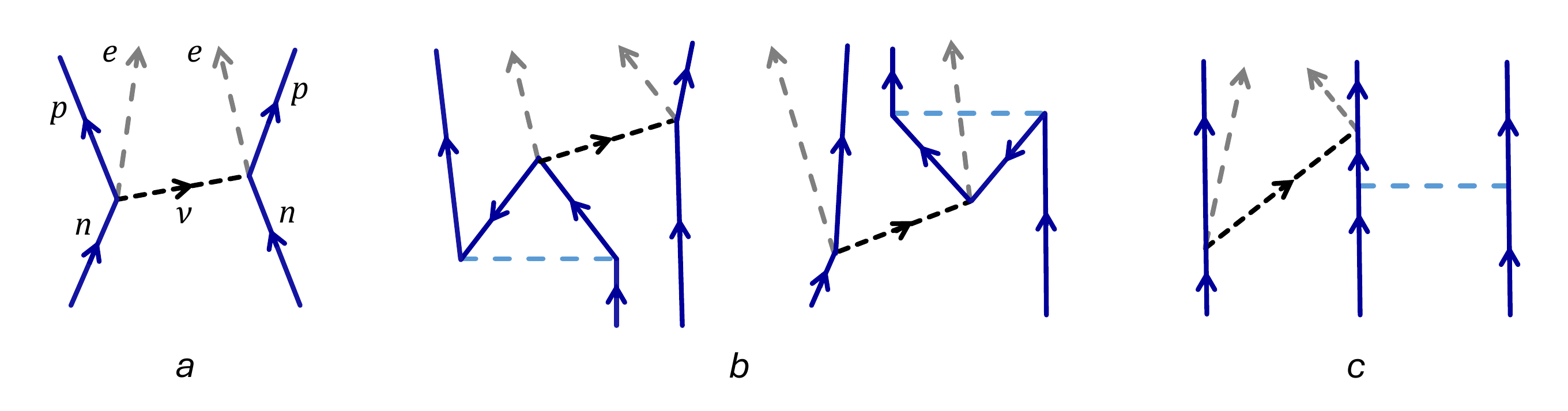}
\vspace{-18pt}
\caption{\label{fig:diagram_0vldvc2bc} \baselineskip=13pt 
Diagrams calculated for $0\nu\beta\beta$ NME. They are the leading-order term (\textit{a}), the vertex corrections (\textit{b}), and the two-body current correction (\textit{c}). The blue solid line shows a nucleon. Those at the lower (upper) ends are the neutrons (protons) included in the initial (final) nuclear ground states. The black dashed line marked $\nu$ in diagram \textit{a} expresses the Majorana neutrino, and the gray dashed line marked $e$ depicts the electron. The horizontal light-blue dashed line in diagrams \textit{b} and \textit{c} expresses the nucleon-nucleon potential.}
\end{minipage}
\end{figure}

We calculated the NME for the $0\nu\beta\beta$ and the $2\nu\beta\beta$ decays of $^{136}$Xe, using the transition operator perturbed in the lowest order by the nucleon-nucleon (NN) potential \cite{Ter25}. The diagrams of the calculated NME terms are illustrated by Fig.~\ref{fig:diagram_0vldvc2bc}. Our two-body current term (diagram \textit{c} of Fig.~\ref{fig:diagram_0vldvc2bc}) is different from that of the chiral effective field theory, in which the $\beta$-decay coupling enters the two-nucleon current \cite{Eng17}. That calculation was performed with the Skyrme interactions (energy density functional) SkM$^\ast$ \cite{Bar82} and SGII \cite{Gia81}, the Coulomb interaction, and the contact like-particle and proton-neutron pairing interactions \cite{Ter20}. The nuclear wave functions were obtained by the QRPA \cite{Ter10}. It was found that the perturbation for the transition operator has a significant effect on  the NME for both  $\beta\beta$ decay modes. Also, in our previous study,  we have  determined the values of   $g_A^\textrm{eff}$ by referring to the half-life calculated with the perturbed NME .
This method is applicable for both the $0\nu\beta\beta$ and $2\nu\beta\beta$ NME on the same footing. It turned out that $g_A^\textrm{eff}$ for the $0\nu\beta\beta$ and $2\nu\beta\beta$ NME are close in the representative calculation. This is indeed remarkable, because it suggests a sort of universality in the value of $g_A^\textrm{eff}$, a notion advanced already in the works of Ref.~\cite{Rho23}.


The purpose of this article is to establish the convergence of the $T_{1/2}^{0\nu}$ calculated using our finding. In other words, the closeness of $g_A^\textrm{eff}$ for the two decay modes is crucial for the prediction of $T_{1/2}^{0\nu}$. This is investigated by variations of the method to calculate $g_A^\textrm{eff}$. In the first half part of this article, we summarize the findings of Ref.~\cite{Ter25}. In the second half part, we examine the convergence of the half-life.

The half-life for the $0\nu\beta\beta$ decay is obtained from
\begin{eqnarray}
\frac{1}{T_{1/2}^{0\nu}} = G_{0\nu} |M_{0\nu}|^2 \left( \frac{\langle m_\nu\rangle}{m_e}\right)^2, \label{eq:halflife_0vbb}
\end{eqnarray}
where our $0\nu\beta\beta$ NME $M_{0\nu}$ is defined by
\begin{eqnarray}
&& M_{0\nu} = M_{0\nu}^\mathrm{GT} -\frac{g_V^2}{g_A^2}M_{0\nu}^\mathrm{F}, \label{eq:M0v_pert}\\ 
\nonumber\\
&& M_{0\nu}^\mathrm{GT} = M_{0\nu}^\mathrm{GT(0)} + M_{0\nu}^\mathrm{GT(vc)} + M_{0\nu}^\mathrm{GT(2bc)}, \label{eq:M0vGT_pert}\\
\nonumber\\[-5pt]
&& M_{0\nu}^\mathrm{F} = M_{0\nu}^\mathrm{F(0)} + M_{0\nu}^\mathrm{F(vc)} + M_{0\nu}^\mathrm{F(2bc)}. \label{eq:M0vF_pert}
\end{eqnarray}
$M_{0\nu}^\mathrm{GT(0)}$, $M_{0\nu}^\mathrm{GT(vc)}$, and $M_{0\nu}^\mathrm{GT(2bc)}$ are the leading, the vertex correction, and the two-body current components, respectively, of the GT NME of the $0\nu\beta\beta$ decay $M_{0\nu}^\mathrm{GT}$. The diagrams \textit{a}, \textit{b}, and \textit{c} in Fig.~\ref{fig:diagram_0vldvc2bc} illustrate $M_{0\nu}^\mathrm{GT(0)}$, $M_{0\nu}^\mathrm{GT(vc)}$, and $M_{0\nu}^\mathrm{GT(2bc)}$, respectively. The analogous definition is used for the Fermi NME of the $0\nu\beta\beta$ decay $M_{0\nu}^\mathrm{F}$, and their diagrams are also given by Fig.~\ref{fig:diagram_0vldvc2bc}. Equation (\ref{eq:halflife_0vbb}) includes the electron mass $m_e$. $G_{0\nu}$ is the phase space factor of the $0\nu\beta\beta$ decay, and this factor is proportional to $g_A^4$ in our convention \cite{Ter25}. The sum of the leading and the correction terms is called the perturbed term; that is $M_{0\nu}^\mathrm{GT}$ and $M_{0\nu}^\mathrm{F}$.

\begin{table}
\caption{\label{tab:NME0v2v} The calculated GT and Fermi NMEs of the $0\nu\beta\beta$ and the $2\nu\beta\beta$ decay modes of $^{136}$Xe with SkM$^\ast$ and SGII. Leading, vc, and 2bc stand for the leading order, the vertex correction, and the two-body current terms, respectively. Sum is called the perturbed NME. The sign is chosen so that the GT leading terms are positive. These results were taken from Ref.~\cite{Ter25}. }
\begin{ruledtabular}
\begin{tabular}{ccccccccc}
& \multicolumn{4}{c}{SkM$^\ast$} & \multicolumn{4}{c}{SGII}\\
\cline{2-5} \cline{6-9}\\[-10pt]
 & \multicolumn{2}{c}{$0\nu\beta\beta$ NME} & \multicolumn{2}{c}{$2\nu\beta\beta$ NME} 
 & \multicolumn{2}{c}{$0\nu\beta\beta$ NME} & \multicolumn{2}{c}{$2\nu\beta\beta$ NME} \\
\cline{2-3} \cline{4-5} \cline{6-7} \cline{8-9}\\[-11pt]
 & GT & Fermi & GT & Fermi & GT & Fermi & GT & Fermi\\
\hline \\[-10pt]
Leading & $\;\;\:$3.095 &      $-$0.467 & $\;\;\:$0.102 &      $-$0.002 & $\;\;\:$2.947 & $-$0.460 & $\;\;\:$0.052 & $-$0.001\\
vc      & $\;\;\:$1.332 &      $-$0.984 & $\;\;\:$0.055 &      $-$0.033 & $\;\;\:$1.456 & $-$0.714 & $\;\;\:$0.037 & $-$0.022\\
2bc     &      $-$2.731 & $\;\;\:$1.758 &      $-$0.192 & $\;\;\:$0.030 & $-$3.497 & $\;\;\:$1.568 & $-$0.067 & $\;\;\:$0.021\\
Sum     & $\;\;\:$1.696 & $\;\;\:$0.307 &      $-$0.035 &      $-$0.005 & $\;\;\:$0.906 & $\;\;\:$0.394 & $\;\;\:$0.022 & $-$0.002\\
\end{tabular}
\end{ruledtabular}
\end{table}

The half-life for the $2\nu\beta\beta$ decay $T_{1/2}^{2\nu}$ can be determined by 
\begin{eqnarray}
\frac{1}{T_{1/2}^{2\nu}} = G_{2\nu} |M_{2\nu}|^2. \label{eq:halflife_2vbb}
\end{eqnarray}
$G_{2\nu}$ is the phase space factor of the $2\nu\beta\beta$ decay, and it is again proportional to $g_A^4$. The $2\nu\beta\beta$ NME $M_{2\nu}$ has the expression corresponding to that of the $M_{0\nu}$;
\begin{eqnarray}
&& M_{2\nu} = M_{2\nu}^\mathrm{GT} -\frac{g_V^2}{g_A^2}M_{2\nu}^\mathrm{F}, \label{eq:M2v_pert}\\ 
\nonumber\\ 
&& M_{2\nu}^\mathrm{GT} = M_{2\nu}^\mathrm{GT(0)} + M_{2\nu}^\mathrm{GT(vc)} + M_{2\nu}^\mathrm{GT(2bc)}, \label{eq:M2vGT_pert}\\ 
\nonumber\\[-5pt]
&& M_{2\nu}^\mathrm{F} = M_{2\nu}^\mathrm{F(0)} + M_{2\nu}^\mathrm{F(vc)} + M_{2\nu}^\mathrm{F(2bc)}. \label{eq:M2vF_pert}
\end{eqnarray}
All calculated NMEs are collected in Table \ref{tab:NME0v2v}. 

The half-life is a function of the GT and Fermi NMEs and $g_A$, so that one can write
\begin{eqnarray}
&&T_{1/2}^{0\nu} = T_{1/2}^{0\nu}(M_{0\nu}^\mathrm{GT}, M_{0\nu}^\mathrm{F},g_A), \\[6pt]
&&T_{1/2}^{2\nu} = T_{1/2}^{2\nu}(M_{2\nu}^\mathrm{GT}, M_{2\nu}^\mathrm{F},g_A). 
\end{eqnarray}
For the $2\nu\beta\beta$ decay, we  define the phenomenological effective $g_A$ denoted by $g_{A,2\nu}^\mathrm{eff}$(ld;exp) by the condition
\begin{eqnarray}
T_{1/2}^{2\nu}(M_{2\nu}^\mathrm{GT(0)}, M_{2\nu}^\mathrm{F(0)},g_{A,2\nu}^\mathrm{eff}(\mathrm{ld;exp})) = T_{1/2}^{2\nu(\mathrm{exp})} \label{eq:gA2veffldexp}, 
\end{eqnarray}
and $g_{A,2\nu}^\mathrm{eff}$(pt;exp) by 
\begin{eqnarray}
T_{1/2}^{2\nu}(M_{2\nu}^\mathrm{GT}, M_{2\nu}^\mathrm{F},g_{A,2\nu}^\mathrm{eff}(\mathrm{pt;exp})) = T_{1/2}^{2\nu(\mathrm{exp})} \label{eq:gA2veffptexp}. 
\end{eqnarray}
$T_{1/2}^{2\nu(\mathrm{exp})}$ is the experimental half-life for the $2\nu\beta\beta$ decay. The first argument of $g_{A,2\nu}^\mathrm{eff}(\mathrm{ld;exp})$, ld, indicates that this $g_A^\textrm{eff}$ is used with the leading GT and Fermi NMEs to calculate the half-life, and that of $g_{A,2\nu}^\mathrm{eff}(\mathrm{pt;exp})$, pt, indicates that the perturbed GT and Fermi NMEs are used. The second argument exp implies that those $g_A^\textrm{eff}$'s are determined to reproduce $T_{1/2}^{2\nu(\mathrm{exp})}$. 

Using the leading and perturbed NME components (the current couplings not included), we can define $g_{A,0\nu}^\mathrm{eff}$(ld;pt) and $g_{A,2\nu}^\mathrm{eff}$(ld;pt) used with the leading NME components  to reproduce the half-life calculated with the perturbed NME components and the bare $g_A$ of 1.267 ($g_A^\mathrm{bare}$). The conditions can be written
\begin{eqnarray}
&&T_{1/2}^{0\nu}(M_{0\nu}^\mathrm{GT(0)}, M_{0\nu}^\mathrm{F(0)},g_{A,0\nu}^\mathrm{eff}(\mathrm{ld};\mathrm{pt})) = T_{1/2}^{0\nu}(M_{0\nu}^\mathrm{GT}, M_{0\nu}^\mathrm{F},g_A^\mathrm{bare}), \label{eq:gA0veffldpt}\\
\nonumber\\[-6pt]
&&T_{1/2}^{2\nu}(M_{2\nu}^\mathrm{GT(0)}, M_{2\nu}^\mathrm{F(0)},g_{A,2\nu}^\mathrm{eff}(\mathrm{ld};\mathrm{pt})) = T_{1/2}^{2\nu}(M_{2\nu}^\mathrm{GT}, M_{2\nu}^\mathrm{F},g_A^\mathrm{bare}). \label{eq:gA2veffldpt} 
\end{eqnarray}
These equations can be rewritten to the forms with only the NME components and the $g_A$'s \cite{Ter25}.

\begin{table}
\caption{\label{tab:gAeff0v2v} Obtained $g_A^\mathrm{eff}$ by different methods, which are defined by the second through the fourth columns. The second column shows the considered decay mode. The third column specifies the GT and Fermi NMEs used with the $g_A^\mathrm{eff}$, and that is either the leading or the perturbed term. The half-life reproduced by the $g_A^\mathrm{eff}$ is specified in the fourth column; see Eqs.~(\ref{eq:gA2veffldexp})$-$(\ref{eq:gA2veffldpt}) and the text. The experimental half-life for the $2\nu\beta\beta$ decay ($T_{1/2}^{2\nu(\mathrm{exp})}$) of $^{136}$Xe is $2.18\times 10^{21}$ y \cite{Bar19}. The fifth and the sixth columns show the values of the $g_A^\mathrm{eff}$ for SkM$^\ast$ and SGII, respectively. these results were taken from Ref.~\cite{Ter25}.}
\begin{ruledtabular}
\begin{tabular}{cccccc}
 Specification & & GT and Fermi & & \multicolumn{2}{c}{$g_A^\mathrm{eff}$}\\
 \cline{5-6}
 of $g_A^\mathrm{eff}$ & Decay & NMEs & Half-life  reproduced  & SkM$^\ast$ & SGII \\
\hline \\[-10pt]
$g_{A,0\nu}^\mathrm{eff}$(ld;pt)   & 
$0\nu\beta\beta$                      & 
Leading                               &
$T_{1/2}^{0\nu}$($M_{0\nu}^\mathrm{GT}, M_{0\nu}^\mathrm{F},g_A^\mathrm{bare}$)       &  
0.796 & 0.454\\[+5pt]
$g_{A,2\nu}^\mathrm{eff}$(ld;pt)                    &
$2\nu\beta\beta$                                    &
Leading &
$T_{1/2}^{2\nu}$($M_{2\nu}^\mathrm{GT}, M_{2\nu}^\mathrm{F},g_A^\mathrm{bare}$) &  
0.696 & 0.847 \\[+5pt]
$g_{A,2\nu}^\mathrm{eff}$(ld;exp)   & $2\nu\beta\beta$ & Leading & $T_{1/2}^{2\nu(\mathrm{exp})}$ & 0.422 & 0.563 \\[+6pt]
$g_{A,2\nu}^\mathrm{eff}$(pt;exp) & $2\nu\beta\beta$ & 
Perturbed & 
$T_{1/2}^{2\nu(\mathrm{exp})}$ & 0.806 & 0.833 \\
\end{tabular}
\end{ruledtabular}
\end{table}
 
We obtained $g_A^\mathrm{eff}$'s by four methods, and they are summarized in Table \ref{tab:gAeff0v2v}. The most remarkable finding in the table is 
\begin{eqnarray}
g_{A,0\nu}^\mathrm{eff}(\mathrm{ld;pt}) \simeq g_{A,2\nu}^\mathrm{eff}(\mathrm{ld;pt}), \ \ (\mathrm{SkM}^\ast). \label{eq:closeness_perturbed_gA}
\end{eqnarray}
This is remarkable because the neutrino momenta $q$ of the $0\nu\beta\beta$ and $2\nu\beta\beta$ decays are quite different. The Majorana neutrino can have infinite $q$ in the neutrino potential because it is a virtual particle. The $g_{A,0\nu}^\mathrm{eff}$ was unknown at all due to the speculation that $g_{A,0\nu}^\textrm{eff}$ might be affected by very high $q$. Our result indicates that $g_{A,2\nu}^\mathrm{eff}(\mathrm{ld;exp})$ and $g_{A,2\nu}^\mathrm{eff}(\mathrm{pt;exp})$ can be used approximately for the $0\nu\beta\beta$ calculation with SkM$^\ast$. The closeness of $g_{A,0\nu}^\mathrm{eff}$ and $g_{A,2\nu}^\mathrm{eff}$ is possible because $g_{A,0\nu}^\mathrm{eff}$ is obtained by the ratio of the perturbed NME to the leading-order NME, and the neutrino potential is included in both the numerator and the denominator \cite{Ter25}. We showed in an analytical discussion \cite{Ter25} that the neutrino potential can be approximately factorized in the numerator and the denominator; thus, the neutrino potential cancels in the ratio. This leads to Eq.~(\ref{eq:closeness_perturbed_gA}). 

The corresponding equation at the leading order is trivially 
\begin{eqnarray}
g_{A,0\nu}^\mathrm{eff}(\mathrm{ld;ld}) = g_{A,2\nu}^\mathrm{eff}(\mathrm{ld;ld}) = g_A^\mathrm{bare}. \label{eq:closeness_bare_gA}
\end{eqnarray}
This and Eq.~(\ref{eq:closeness_perturbed_gA}) imply that the ratio of $g_{A,0\nu}^\mathrm{eff}$ to $g_{A,2\nu}^\mathrm{eff}$ is close to the convergence at the first-order perturbation. The infinite-order $g_{A,2\nu}^\mathrm{eff}$ is equal to  the phenomenological $g_{A,2\nu}$ to reproduce $T_{1/2}^{2\nu(\mathrm{exp})}$. We can use this for the nonperturbative calculation of $M_{0\nu}$. 

The first step of new discussion is to consider implications of the three $g_{A,2\nu}^\textrm{eff}$ in Table \ref{tab:gAeff0v2v}. $T_{1/2}^{2\nu(\textrm{exp})}$ is $2.18\times 10^{21}$ y \cite{Bar19}, and $T_{1/2}^{2\nu}(M_{2\nu}^\textrm{GT},M_{2\nu}^\textrm{F},g_A^\textrm{bare})$ is $0.26\times 10^{21}$ y (SkM$^\ast$). Thus, we have the relations  
\begin{eqnarray}
&& T_{1/2}^{2\nu(\textrm{exp})} > 
T_{1/2}^{2\nu}(M_{2\nu}^\textrm{GT},M_{2\nu}^\textrm{F},g_A^\textrm{bare}), \label{eq:T2vexp_T2vpert}\\
&& g_{A,2\nu}^\mathrm{eff}\textrm{(ld;exp)} < g_{A,2\nu}^\mathrm{eff}\textrm{(ld;pt)}.
\end{eqnarray}
The SGII calculation has the same relations. These relations can be understood from Eq.~(\ref{eq:halflife_2vbb}), that is, the smaller $g_{A,2\nu}^\textrm{eff}$ is necessary to reproduce the larger $T_{1/2}^{2\nu}$ if the same NME components are used; note that   
\begin{eqnarray}
G_{2\nu}|M_{2\nu}|^2 \propto | g_A^2 M_{2\nu}^\textrm{GT}-g_V^2 M_{2\nu}^\textrm{F}|^2.
\end{eqnarray}
$T_{1/2}^{2\nu(\textrm{exp})}$ includes more perturbation effects of the transition operator than our calculations. Therefore, it is seen that the smaller $g_{A,2\nu}^\textrm{eff}$ reflects the larger perturbation effects. 
 Comparing $g_{A,2\nu}^\mathrm{eff}\textrm{(ld;exp)}$ and $g_{A,2\nu}^\mathrm{eff}\textrm{(pt;exp)}$, we see that the NME components with the larger perturbation effects yield the larger $g_{A,2\nu}^\textrm{eff}$ if the same half-life is reproduced. Because of the perturbation effects carried by the perturbed NME components, $g_{A,2\nu}^\textrm{eff}$ does not need to have strong perturbation effects. Dynamically, the magnitude relations of the $g_A^\textrm{eff}$ and the half-life can be understood in terms of the perturbation effects of the transition operator. 
Likewise, the relatively close relation of $g_{A,2\nu}^\mathrm{eff}\textrm{(ld;pt)}$ and $g_{A,2\nu}^\mathrm{eff}\textrm{(pt;exp)}$ can also be understood. The use of the perturbed NME components makes $g_{A,2\nu}^\textrm{eff}$ larger, but the more perturbed $T_{1/2}^{2\nu}$ needs smaller $g_{A,2\nu}^\textrm{eff}$. Thus, $g_{A,2\nu}^\mathrm{eff}\textrm{(ld;pt)}$ $\simeq$  $g_{A,2\nu}^\mathrm{eff}\textrm{(pt;exp)}$ can happen. 
The appropriate $g_A^\textrm{eff}$ depends on the calculation method of the NME. It is close to one in our calculation with the lowest-order correction to the transition operator.

Now, we test the estimated $g_{A,0\nu}^\mathrm{eff}$ defined by
\begin{eqnarray}
&& g_{A,0\nu}^\mathrm{eff}\mathrm{(ld;est)} = g_{A,2\nu}^\mathrm{eff}\mathrm{(ld;exp)}\frac{g_{A,0\nu}^\mathrm{eff}\mathrm{(ld;pt)} }{g_{A,2\nu}^\mathrm{eff}\mathrm{(ld;pt)} }, \label{eq:gA0veffldest}
\end{eqnarray}
for the calculation of $T_{1/2}^{0\nu}$. If the perturbation calculation is close to the exact one, the right-hand side is approximately equal to $g_{A,0\nu}^\mathrm{eff}\mathrm{(ld;pt)}$, which is close to the exact one. Equation (\ref{eq:gA0veffldest}) has the higer-order effects than the leading-order calculation because of the correction factor (the fraction part). This structure is the reason for introducing this correction factor. 
We also test, with the same correction factor, 
\begin{eqnarray}
&& g_{A,0\nu}^\mathrm{eff}\mathrm{(pt;est)} = g_{A,2\nu}^\mathrm{eff}\mathrm{(pt;exp)}\frac{g_{A,0\nu}^\mathrm{eff}\mathrm{(ld;pt)} }{g_{A,2\nu}^\mathrm{eff}\mathrm{(ld;pt)} }.
\end{eqnarray}
Below, we use these $g_{A,0\nu}^\mathrm{eff}$ not only for the SkM$^\ast$ but also for the SGII calculation. The values of these $g_{A,0\nu}^\mathrm{eff}$ are shown in Table \ref{tab:gAeff0vest}. The correction factor is 1.14 for SkM$^\ast$ and 0.54 for SGII. 

\begin{table}
\caption{\label{tab:gAeff0vest} Estimated $g_{A,0\nu}^\mathrm{eff}$ for SkM$^\ast$ and SGII calculations. The second column shows the GT and Fermi NMEs used with the $g_{A,0\nu}^\mathrm{eff}$.}
\begin{ruledtabular}
\begin{tabular}{cccc}
Specification of   & GT and Fermi & \multicolumn{2}{c}{$g_A^\mathrm{eff}$}\\
 \cline{3-4}
$g_A^\mathrm{eff}$ & NMEs         &  SkM$^\ast$ & SGII \\
\hline \\[-10pt]
$g_{A,0\nu}^\mathrm{eff}$(ld;est) & 
Leading                           &
0.482                             &  
0.302                             \\[+5pt]
$g_{A,0\nu}^\mathrm{eff}$(pt;est) &
Perturbed                         &
0.921                             &
0.447                              
\end{tabular}
\end{ruledtabular}
\end{table}

\begin{figure}[t]
\begin{minipage}[t]{0.9\linewidth}
\includegraphics[width=0.6\columnwidth]{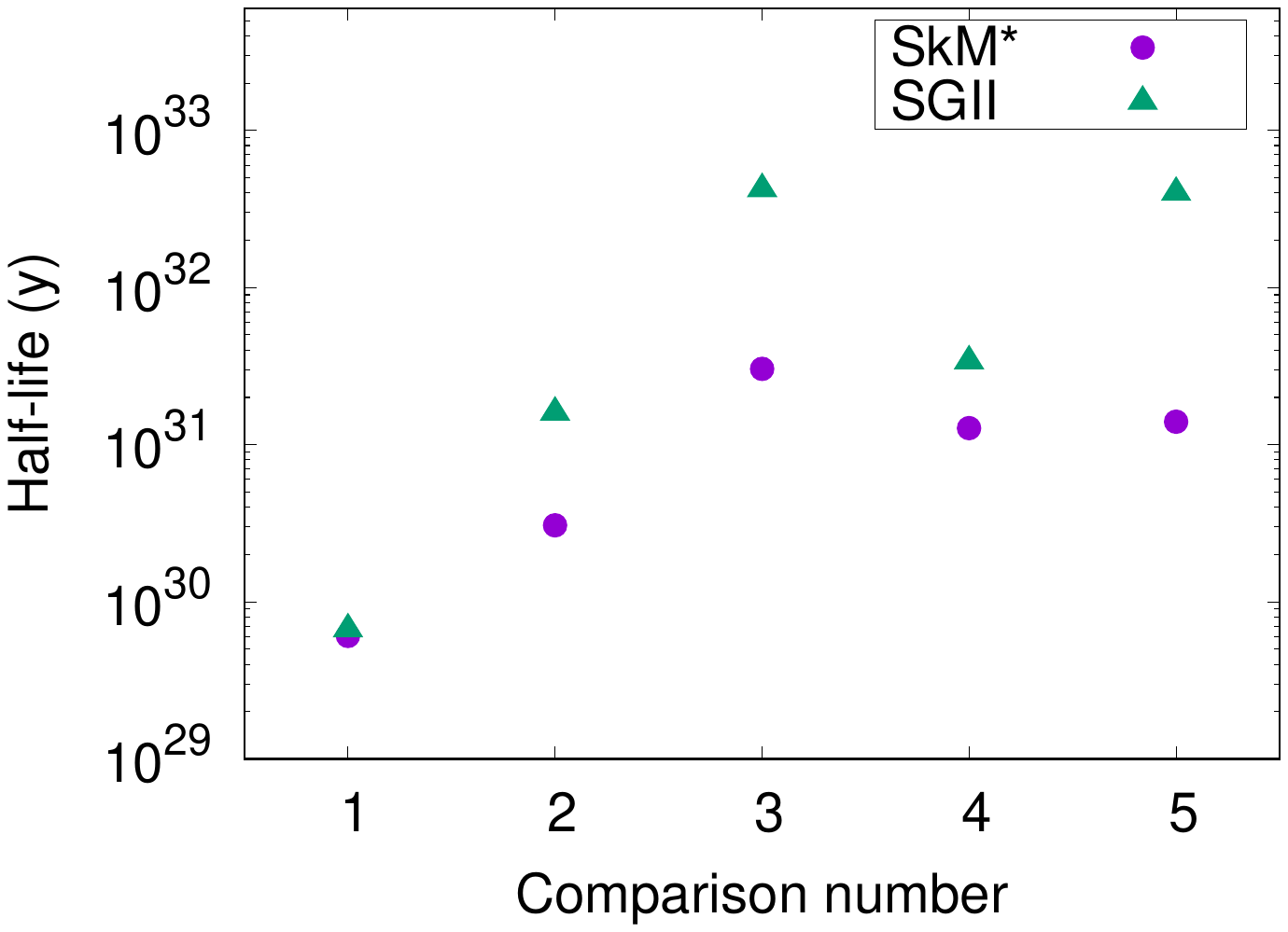}
\vspace{-10pt}
\caption{\label{fig:hl_0vbb} \baselineskip=13pt 
Illustration of calculated $T_{1/2}^{0\nu}$ with assumed $\langle m_\nu\rangle$ = 1 meV given by Table \ref{tab:hl_0vbb}.}
\end{minipage}
\end{figure}

We obtained ten values of the $0\nu\beta\beta$ half-lives with different interactions, the leading and the perturbed NME components, and different $g_{A,0\nu}^\mathrm{eff}$'s. These half-lives are summarized in Table \ref{tab:hl_0vbb} and illustrated in Fig.~\ref{fig:hl_0vbb}. 
Now, we discuss which half-life is most reliable.
In this article, all half-lives for the $0\nu\beta\beta$ decay are obtained assuming arbitrarily $\langle m_\nu\rangle$ = 1 meV. If $\langle m_\nu\rangle$ = 10 meV, the half-lives are two orders of magnitude shorter.
The calculated half-lives are distributed from 6$\times$10$^{29}$ y to 4$\times$10$^{32}$ y. 
The shortest is that with $g_A^\mathrm{bare}$ and the leading NME components; both the SkM$^\ast$ and SGII calculations have the similar values. 
It has been shown \cite{Ter25} that the lowest-order perturbation for the transition operator gives a significant effect on the NME. Thus, the shortest half-lives of the lowest-order calculations can be excluded. 

\begin{table}
\caption{\label{tab:hl_0vbb} Calculated $T_{1/2}^{0\nu}$ of $^{136}$Xe with assumed $\langle m_\nu\rangle$ = 1 meV for the two interactions. Five methods (comparison numbers 1$-$5) were used to obtain $T_{1/2}^{0\nu}$. The definitions of those methods are shown in the second and the third columns. Those $T_{1/2}^{0\nu}$ are also shown in Fig.~\ref{fig:hl_0vbb}.  }
\begin{ruledtabular}
\begin{tabular}{cccrr}
 & & & \multicolumn{2}{c}{Half-life (10$^{29}$ y)} \\[-3pt]
 & & & \multicolumn{2}{c}{with $\langle m_\nu\rangle$ = 1 meV} \\
\cline{4-5}\\[-11pt]
Comparison  &       & GT and Fermi &            & \\[-3pt]
number      & $g_A$ &  NMEs          & SkM$^\ast$ & SGII \\
\hline \\[-10pt]
1 & $g_A^\mathrm{bare}$                & Leading   &  6  & 7    \\
2 & $g_A^\mathrm{bare}$                & Perturbed & 31  & 159  \\
3 & $g_{A,0\nu}^\mathrm{eff}$(ld;pt)   & Perturbed & 304 & 4210 \\
4 & $g_{A,0\nu}^\mathrm{eff}$(ld;est)  & Leading   & 127 & 337  \\
5 & $g_{A,0\nu}^\mathrm{eff}$(pt;est)  & Perturbed & 140 & 3990
\end{tabular}
\end{ruledtabular}
\end{table}

As we discussed in our previous article \cite{Ter25}, SGII has a too strong attractive effect on the binding energy, and it is important to take this into account because our correction terms depend on the NN potential explicitly. The stronger interaction causes more perturbation, which causes longer half-lives. 
This is reflected in the relation of the SGII and the SkM$^\ast$ calculations; see Fig.~\ref{fig:hl_0vbb}. Therefore, the SkM$^\ast$ calculation is more reliable than the SGII calculation. 

For this reason we shall concentrate on the analysis of the results obtained with the use of the SkM$^\ast$ interaction. This is important to have in mind because this interaction is the same which have been used in other seemingly unrelated scenario \cite{Rho25}.
We used a few approximations in our perturbed-NME calculations. Above all, the higher-order corrections are ignored. Thus, it is inferred that more perturbation is necessary on top of the SkM$^\ast$ calculation of Comparison 2 (see Table \ref{tab:hl_0vbb}). We assume that the $g_A^\mathrm{bare}$ with the true GT and Fermi components can give the correct half-life. The half-lives of Comparisons 3$-$5 are obtained with more perturbation than that of Comparison 2, and they are in the same order of 10$^{31}$ y. The experimental $2\nu\beta\beta$ half-life used for Comparisons 4 and 5 is a non-perturbative value. This stability against the difference in the method is a positive indication of its correctness. The closeness of the results of Comparisons 4 and 5 implies the possibility that the half-life of this estimation method is converged because of the non-perturbative calibration if the effects of the residual interaction are small. Our most reliable estimation of the $0\nu\beta\beta$ half-life is concluded to be those of Comparisons 3$–$5 (SkM$^\ast$), that is, (1.3$–$3.0)$\times$10$^{31}$ y with assumed $\langle m_\nu\rangle$ = 1 meV.


Up to now, the $g_{A,0\nu}^\mathrm{eff}\approx 1$ obtained by our method with the perturbed NME components and SkM$^{\ast}$ has an interesting coincidence with other studies, namely;
\begin{itemize}
\item{The present estimate $g_{A,0\nu}^\mathrm{eff} \approx 1$ is in concordance with the value reported  very recently  by M. Rho \cite{Rho25}. 
In Rho's view, his analysis, developed in a nuclear matter context, could eventually be supported by the measurement of super-allowed GT $\beta$ decays in heavy mass nuclei with densities close to the nuclear matter density. By performing an  
analysis of the GSI  measurements of super-allowed  GT $\beta$ decay in $^{100}$Sn \cite{Hin12} he concluded that $g_{A}^\mathrm{eff} \approx 1$.}

\item{Also, the same magnitude of $g_{A}^\mathrm{eff}$ was reported in Rho' works \cite{Rho23,Rho25}  in a different context and for the same interaction SkM$^{\ast}$, this time interpreted in terms of a hidden symmetry .}
\item{Values of $g_A^\mathrm{eff}$ for the $\beta$ decay of light-mass nuclei are also similar \cite{Suh17c} to the value reported in our present work for the decay of the heavy-mass nucleus $^{136}$Xe.}
\end{itemize}
In summary, we have calculated the half-life of $^{136}$Xe for the $0\nu\beta\beta$ decay, including the perturbation for the transition operator. Our GT and Fermi NMEs were derived by applying the second-order Rayleigh-Schr\"{o}dinger perturbation theory and choosing the terms corresponding to the diagrams. A new estimation of $g_A^\textrm{eff}$ for the $0\nu\beta\beta$ decay was proposed using the experimental half-life for the $2\nu\beta\beta$ decay. This is a non-perturbative modification. The key that opened the door of our new method is the finding that $g_{A,0\nu}^\mathrm{eff}$ and $g_{A,2\nu}^\mathrm{eff}$ obtained perturbatively are close to each other if the appropriate interaction is used. 

We have obtained ten estimates of the $0\nu\beta\beta$ half-life with the different combinations of the interaction, $g_{A,0\nu}^\mathrm{eff}$, and the NME components. We have selected the ones which are stable with respect to the methodological variation. These half-lives include much more  perturbation effects than that of the leading-order calculation. We predict the half-life for the $0\nu\beta\beta$ decay of $^{136}$Xe to be (1.3$-$3.0)$\times$10$^{31}$ y ($\langle m_\nu\rangle$ = 1 meV). This is much longer than (3.0$-$30.0)$\times$10$^{29}$ y shown by the compilation of the calculations of different groups in Ref.~\cite{Eng17}.
We also note the $0\nu\beta\beta$ $T_{1/2}^{0\nu}$ values obtained from the NMEs of more recent references and $\langle m_\nu \rangle$ = 1 meV. Reference \cite{Cor20b} presents $M_{0\nu}$ of 2.6, and the corresponding $T_{1/2}^{0\nu}$ of 0.1$\times 10^{31}$ y is obtained using $g_A^\textrm{bare}$. This reference also reports $M_{0\nu}$ = 1.15 with a quenching factor $q$ of 0.61, from which $T_{1/2}^{0\nu}$ = 3.8$\times 10^{31}$ y is derived. The authors of Ref.~\cite{Bra22} obtained $M_{0\nu}$ = $0.57^{+0.56}_{-0.34}$ with $q$ = 0.65 and $M_{0\nu}$ = $0.73^{+69}_{-44}$ with $q$ = 0.42. 
The corresponding $T_{1/2}^{0\nu}$ is equal to $2.1^{+11.0}_{-1.6}\times 10^{31}$ y and $1.3^{+7.0}_{-1.0}\times 10^{31}$ y, respectively. The NME in Table IV of  Ref.~\cite{Jok23} is 0.98$-$3.11 with $q$ = 0.42$-$0.72. If the lower $q$ yields the larger NME, $T_{1/2}^{0\nu}$ = (2.3$-$2.7)$\times 10^{31}$ y is obtained. Thus, our half-life has an overlap with some of the other recent calculations.

We have three findings to support our new idea of using $g_{A,2\nu}^\mathrm{eff}$ for $g_{A,0\nu}^\mathrm{eff}$. The first is the convergence of $g_{A,0\nu}^\mathrm{eff}/g_{A,2\nu}^\mathrm{eff} \approx 1$ with respect to the perturbation for the transition operator. This is shown by Eqs.~(\ref{eq:closeness_perturbed_gA}) and (\ref{eq:closeness_bare_gA}). The second is the analytical discussion on the question of why the neutrino potential does not significantly change $g_{A,0\nu}^\mathrm{eff}$ \cite{Ter25}. The basic reason is that  $g_A^\mathrm{eff}$ is given by a ratio of the perturbed NME component to the leading-order one. Because of this, the effect of the neutrino potential on the $0\nu\beta\beta$ NME components  approximately cancels in $g_{A,0\nu}^\mathrm{eff}$. This was confirmed by analyzing the numerically obtained results. 
The third is the stability of the calculated half-life with respect to the methodological variation shown by Fig.~\ref{fig:hl_0vbb}. Therefore, a breakthrough has been made for $g_{A,0\nu}^\mathrm{eff}$ that has been unknown for many years.

\begin{acknowledgments}  
This study was supported by the Czech Science Foundation (GA\v{C}R), project No. 24-10180S. The computation for this study was performed by Karolina (OPEN-33-78), IT4Innovations supported by the Ministry of Education, Youth and Sports of the Czech Republic through the e-INFRA CZ (ID:90254); the computers of MetaCentrum provided by the e-INFRA CZ project (ID:90254), supported by the Ministry of Education, Youth and Sports of the Czech Republic; and Yukawa-21 at Yukawa Institute for Theoretical Physics, Kyoto University. O.C acknowledges the PIP 2081 (CONICET) and the PICT  40492 (ANPCyT). 
\end{acknowledgments}


\bibliography{version2.6}
\end{document}